\documentstyle[prl,twocolumn,aps,epsf,ifthen,epsfig]{revtex}

\newcommand{\nc}{\newcommand}
\nc{\pt}{p_{\rm T}}
\nc{\qt}{q_{\rm T}}
\nc{\Mt}{M_{\rm T}}
\nc{\se}{\section}
\nc{\suse}{\subsection}
\nc{\beq}[1]{\begin{equation}\label{#1}}
\nc{\eeq}{\end{equation}}
\nc{\bea}[1]{\begin{eqnarray}\label{#1}}
\nc{\eea}{\end{eqnarray}}
\nc{\bce}{\begin{center}}
\nc{\ece}{\end{center}}
\nc{\bit}{\begin{itemize}}
\nc{\eit}{\end{itemize}}
\nc{\bmp}{\begin{minipage}}
\nc{\emp}{\end{minipage}}

\nc{\la}{\langle}       
\nc{\lla}{\left \langle}
\nc{\ra}{\rangle}       
\nc{\rra}{\right \rangle}

\newcommand{\prep}[3]{Phys. Rep. {\bf #1}, #3 (#2)}
\newcommand{\prevc}[3]{Phys. Rev. {\bf C#1}, #3 (#2)}
\newcommand{\prevd}[3]{Phys. Rev. {\bf D#1}, #3 (#2)}
\newcommand{\prevl}[3]{Phys. Rev. Lett.\ {\bf #1}, #3 (#2)}

\newcommand{\plb}[3]{Phys. Lett. {\bf B#1}, #3 (#2)}
\newcommand{\npa}[3]{Nucl. Phys. {\bf A#1}, #3 (#2)}

\begin{document}

\twocolumn[\hsize\textwidth\columnwidth\hsize\csname @twocolumnfalse\endcsname

\title{Spectroscopy of resonance decays in high-energy heavy-ion experiments}
\author{Peter F.~Kolb and Madappa Prakash} 
\address{Department of Physics and Astronomy, 
         SUNY, Stony Brook, NY 11794-3800, USA } 
\date{April 23, 2003}

\maketitle


\begin{abstract}
Invariant mass distributions of the hadronic decay products from
resonances formed in relativistic heavy ion collision (RHIC)
experiments are investigated with a view to disentangle the effects of
thermal motion and the phase space of decay products from those of
intrinsic changes in the structure of resonances at the freeze-out
conditions.  Analytic results of peak mass shifts for the cases of
both equal and unequal mass decay products are derived.  The shift is
expressed in terms of the peak mass and width of the vacuum or
medium-modified spectral functions and temperature.  Examples of
expected shifts in meson ({\it e.g.}, $\rho,~\omega,~{\rm
and}~\sigma$) and baryon ({\it e.g.}, $\Delta$) resonances that are
helpful to interpret recent RHIC measurements at BNL are provided.
Although significant downward mass shifts are caused by widened widths
of the $\rho-$meson in medium, a downward shift of at least 50 MeV in
its intrinsic mass is required to account for the reported downward
shift of 60-70 MeV in the peak of the $\rho-$invariant mass
distribution. An observed downward shift from the vacuum peak value of
the $\Delta$ distinctively signals a significant downward shift in its
intrinsic peak mass, since unlike for the $\rho-$meson, phase space
functions produce an upward shift for the $\Delta-$ isobar. 
\\

PACS numbers: 25.75-q, 25.75.Dw, 13.25-k, 13.30.Eg
\end{abstract}
\vspace{0.4cm}
]
\begin{narrowtext}
\newpage

\se{Introduction}

Decay products from unstable resonances whose lifetimes are smaller
than the duration of $\sim 10-15$ fm/c of a relativistic heavy ion
collision (RHIC) give valuable insights into (a) the ambient
conditions of strongly interacting matter (such as its temperature and
chemical potential if matter is equilibrated), and (b) the influence
of the hot and dense medium on their intrinsic properties (such as
their masses and widths).
Through their electromagnetic decays, vector meson 
resonances like $\rho,~\omega,~a_1,$~etc... enable a direct view
into hot hadronic matter, since dileptons and photons escape without
further interactions with matter (for recent accounts, 
see Refs.~\cite{HBEK02,Gale02,RW00}).
In contrast, hadronic decays which occur when the medium is dense and
strongly interacting contribute to thermalize matter, since the decay
products undergo substantial rescattering before escaping. Hence such
hadronic decay products cannot be used to infer the properties of the
parent resonances.

An intriguing exception is the case in which resonance decays occur
close to the thermal freeze-out of hadrons.  If the decay
products do not further interact with the hadronic environment, and
their total four-momentum $q$ can be measured (this requires that the
particles in the final state can be identified and their momenta
measured), a resonance bump should appear in the invariant mass plots
in $M = \sqrt s = {\sqrt {q^2}}$.
Such reconstructed resonances constitute spectroscopic probes of the
freeze-out stage of RHIC experiments.  Where possible, a comparison of
invariant mass plots from $e^+e^-$ or $pp$ collisions, with those from
nucleus-nucleus collisions at the same c.m. energy enables us
to establish the extent to which medium effects leave
their imprint on resonances. As we demonstrate, shifts in the peak
masses for the cases of $\rho-$meson and $\Delta-$isobar decays are
distinctively different at the expected freeze-out conditions of 
RHIC collisions.    

Thanks to the growing multiplicities of final state hadrons in RHIC
experiments, detailed spectral information is beginning to become
available.  Examples from Au+Au collisions at BNL include the
reconstruction of (a) $K^\star(892)^0$ mesons \cite{STAR130K*} and
$\phi$-mesons \cite{STAR130phi} at c.m. energy $\sqrt{s_{\rm NN}}=
130$ GeV, (b) $\rho$-decays~\cite{Fachini02} at $\sqrt{s_{\rm NN}} =
200$ GeV (preliminary results), and (c) $\Lambda,~\Xi^-, {\rm and}~
\Omega^-$ decays~\cite{Adams02} at $\sqrt{s_{\rm NN}} = 130$ GeV.
Among the results available to date, the $60-70$ MeV downward shift of
the $\rho-$meson peak from its vacuum value of 776 MeV is interesting
both because such a shift, if confirmed, implies the presence of
considerable hadronic interactions at freeze-out and because it
provides the opportunity to pin down in-medium properties of mesons
that are susceptible to effects of partial chiral symmetry restoration.

Our objective in this work is to point out how to benefit from such
detailed information. Toward this goal, we study invariant mass
distributions from the hadronic decay of unstable resonances by
exploring the combined effects of phase space functions associated
with thermal motion and the decay products, and possible intrinsic
changes in the properties of resonances in a hadronic heat bath.
While phase space modifications can be evaluated without
approximation, those induced by intrinsic changes of resonance
structure require model dependent considerations. Ongoing debates
about medium modifications of spectral functions can be found, for
example, in Refs. \cite{EBEK02,KKW97,RCW96,SYZ96,Pisarski95,BR91}.  In
order to highlight the basic trends of in-medium effects on resonance
properties, we utilize results of model calculations in vogue in
conjunction with the inevitable phase space functions to show how the
invariant mass distributions of resonance decay products are modified
from their vacuum counterparts. 

\section{Invariant mass distributions from the decay of unstable
         resonances }

In thermalized matter, the number of events in which a
resonance $R$ is produced and subsequently decays into a final state
$f$ is given by (see, for example, Ref.~\cite{Weldon93})
\beq{equ:distribution} \frac {dN_f}{d^4x d^4q} = 
\frac{dN_R}{d^3 x d^3 q} \frac{ (-{\rm Im}~\Pi)(M_R/ \pi)} {
(M^2 - M_R^2)^2 + ({\rm Im}~\Pi)^2} 2 \Gamma_{R \rightarrow f}^{\rm
vac} \,.  
\eeq 
Above, $q$ is the total four-momentum of the decay products,
$M^2=s=q^2$ is the invariant mass squared, $dN_R/d^3xd^3q$ is the
thermal phase space (Bose-Einstein or Fermi-Dirac) distribution of the
off-shell unstable resonance particle, $M_R$ is the mass of the
resonance at its peak, and $\Gamma_{R \rightarrow f}^{\rm vac}$ is the
decay width in vacuum.  The proper self-energy $\Pi$ describes the
influence of the heat bath on the properties of the resonance;
explicitly, 
\beq{equ:self-energy} {\rm Im~\Pi} = -M_R \Gamma_{tot}\,, 
\qquad M_R^2 = (M_R^{vac})^2 +  {\rm Re~\Pi} \,.
\eeq 
In general, the resonance mass $M_R$ and the width $\Gamma_{tot}$
depend on the temperature $T$, chemical potential $\mu$, and
interactions between the various constituents of the heat bath.
Additionally, both $M_R$ and $\Gamma_{tot}$ depend on the momentum of
the resonance~\cite{EBEK02}.  Eq.~(\ref{equ:distribution}) applies
strictly to the case in which the hadronic decay products do not undergo
further interactions within the medium and freely escape to the
detectors. Its applicability is therefore contingent on rescattering
effects being small at the decoupling or freeze-out conditions.

The thermal distribution of a resonance particle is
\beq{equ:thermaldist}
   \frac{dN_R}{d^3x d^3q}
   =
   \frac{(2 J+1)}{(2 \pi)^3}
   \frac{1}{e^{\, \beta (E-\mu)} -\sigma}
\,,
\eeq
where $J$ denotes the spin of the resonance, $1/\beta =T$ is the
temperature, $\sigma = 1$ for bosons and $-1$ for fermions, and
$E=M_{\rm T} \cosh y$ is the energy of the resonance with rapidity $y$
and transverse mass $M_{\rm T} = \sqrt{M^2+\qt^2}$\,, in the fluid
rest frame.

In what follows, we utilize Eqs.~(\ref{equ:distribution}),
(\ref{equ:self-energy}), and  
(\ref{equ:thermaldist}) to study the invariant mass distribution of
the decay products stemming from resonances in a mixture of
thermalized hadrons. We begin by first considering $\rho \rightarrow
\pi\pi$ decays. In vacuum, we take the decay width of the $\rho$ 
(the decay occurs with an orbital angular momentum of $\ell =1$) to be
\beq{equ:momspace} 
\Gamma_{R \rightarrow f}^{\rm vac} = \Gamma_R
\left( \frac {k}{k_0}\right)^{2\ell +1}  
=  \Gamma_R 
\left(\frac{M^2 - 4 \, m_\pi^2 } {M_R^2 - 4 \, m_\pi^2 } \right)^{3/2} \,, 
\eeq 
for pion c.m. momenta $k$ not too far in excess of $k_0$, the value at
the peak of the resonance.  The width grows as $k^{(2\ell + 1)}$ due
to the penetration factor associated with a partial wave of orbital
angular momentum $\ell$.  For $k\gg k_0$, the monotonically increasing
$k-$dependence in this equation should, however, be tamed as noted
{\it e.g.}, in Ref.~\cite{BGMRS88}.  Alternative forms for $\Gamma_{R
\rightarrow f}^{\rm vac}$ have been used in the literature (see, for
example, Ref.~\cite{BRW86}).  Since our primary interest is limited to
the peak region, Eq.~({\ref{equ:momspace}}) provides an adequate
representation insofar as these different parametrizations all agree
near the Breit-Wigner peak.
The invariant mass distribution is obtained from
Eq.~(\ref{equ:distribution}) by integrating out the longitudinal and
transverse momenta of the resonance, {\it i.e.},
\beq{equ:invmassdist}
\frac{dN_f}{d^4x dM}
=
\pi\,M \, \int \frac{dN_f}{d^4x d^4q} dy \, \qt d\qt \,.
\eeq
Equations (\ref{equ:distribution}) through (\ref{equ:invmassdist})
enable us to explore the combined effects of intrinsic changes in the
properties of the resonance in a heat bath (given by the real and
imaginary parts of the self-energy $\Pi$) and phase space functions
associated with thermal motion and decay products.  In order to
highlight the effects of each, we consider in turn the cases in which
the resonance mass and width are (a) not affected by the heat bath,
and (b) influenced by interactions with the other constituents in the
heat bath.

\suse{ Shifts induced by the phase spaces of thermal motion
       and decay products }               

In this case, ${\rm Re}~\Pi$ is set to zero ({\it i.e.}, no intrinsic
mass shift) and $\Gamma_R$ is taken at its vacuum value. This allows
us to cast Eq.~(\ref{equ:invmassdist}) as
\bea{equ:noshifts}
\frac{dN_f}{d^4xdM} &=&
      \frac{2 M M_R^2 \Gamma_R  \Gamma_{R\rightarrow f}^{\rm vac}}
           {(M^2-M_R^2)^2+(M_R \Gamma_R)^2}
       I(M,\,\mu,\,T) \,;  \nonumber  \\    
 I(M,\,\mu,\,T) &=&\frac{2J+1}{(2 \pi)^3} 
  \int dy \, \qt d\qt \;\frac{1}{e^{(E-\mu)/T}-\sigma}\,.
\eea
For $M/T \gg 1$, it is advantageous to
express $I$ above (after performing integrations over $q_T$ and
$y$) as
\beq{equ:I2}
  I(M,\,\mu,\,T) 
  =\frac{2J+1}{(2 \pi)^3}  
  2 M T \sum_{j=1}^{\infty} (\pm)^{j-1} \frac{e^{j \beta \mu}}{j} 
{\rm K}_1\left(\frac{j M}{T}\right) \,,
\eeq
where the $+(-)$ sign refers to bosons (fermions) and ${\rm K}_1$ is
the modified Bessel function. 
For invariant masses $M \simeq M_R$, where $M_R
\simeq 776$ MeV, and temperatures in the range $100 \leq T/{\rm MeV}
\leq 175$, we can retain only the $j=1$ term in the sum above and
employ the asymptotic expansion $K_\nu(x) \simeq (\pi/2x)^{1/2}
\exp(-x)$ in order to gain a qualitative insight into the shifts
caused by phase space alone.  Collecting the $M-$dependences from
Eqs.~(\ref{equ:invmassdist}), (\ref{equ:noshifts}), and (\ref{equ:I2})
together, we arrive at
\bea{equ:qual}
\frac{dN_f}{d^4xdM} & \propto & 
\frac{1 } {(M^2-M_R^2)^2+(M_R \Gamma_R)^2} \times (M^2-4m_\pi^2)^{3/2}
\nonumber \\ 
& \times & (MT)^{3/2} \exp\left(\frac{\mu-M}{T}\right) \,,
\eea
where the factors omitted, including those that depend on $M_R$ and
$\Gamma_R$, contribute only to the normalization.  The result in
Eq.~(\ref{equ:qual}) has the appealing physical interpretation that
the invariant mass distribution is given by the simple product
``Breit-Wigner probability of resonance formation'' $\times$
``Available phase space for decay products'' $\times$ ``Density of
Boltzmann-like `particles' of mass $M$.'' (Such simplicity is lost
when effects of interactions and Bose-Einstein enhancement are
included.)  Note that while the Boltzmann factor $\exp(-M/T)$ tends to
pull the peak position of the Breit-Wigner at $M_R$ to lower values of
$M$ (the lower the $T$, the larger is the downward shift), the
$M^{4.5}$ power law dependence arising from the phase space of decay
products and the pre-exponential factor in the density tend to pull
the peak position to $M > M_R$.  The net result from
Eq.~(\ref{equ:noshifts}) is shown in Fig.~\ref{fig:thermalshifts},
where we have used $M_R=775.9$~MeV and $\Gamma_R= 147.9$~MeV inferred
from $\rho-$ production in $e^+e^-$ collisions \cite{PDG02} which are
free of uncertainties associated with $\rho-$ production in hadronic
collisions~{\footnote{The resonant part of the $\rho-$mass spectrum in
hadronic collisions is well fit by a Breit-Wigner form with
$M_R(\rho)=766.5$ MeV and $\Gamma_R(\rho)=150.2$ MeV
\cite{PDG02}. Relative to the spectrum observed in $e^+e^-$
collisions, there is $\sim 10$ MeV downward shift in the peak mass,
which is commonly attributed to effects of final state interactions.
While it is tempting to use this downward shift to advantage, it is,
as yet, unclear if the effects of final state interactions in
heavy-ion collisions are the same as in elementary hadronic
collisions. Use of such an intrinsic shift is, however, 
straightforward in the theoretical formalism outlined in Secs. II
through IV.}}.
The results shown refer to the case of $\mu=0$; nonzero
values of $\mu$ mainly affect the normalization of the distributions,
but not their peak positions or shapes.  It is clear that the combined
effects of thermal and decay product phase spaces are small, and at
best produce a downward shift of about $10$ MeV at the lowest
temperature in the range $100 \leq T/{\rm MeV} \leq 175$.

 \begin{figure}[t,b,p]
 \epsfig{file=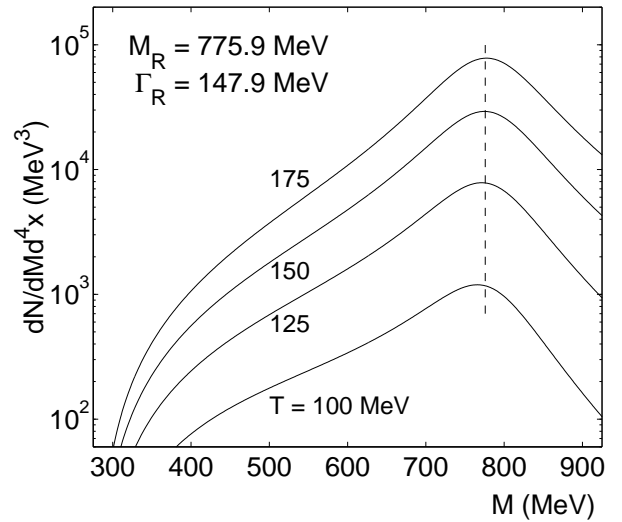,width=8cm,clip=}\\
 \caption{$\rho-$meson invariant mass distributions in a heat bath at
          the indicated temperatures $T$.  Results exhibit effects of
          thermal and decay product phase spaces alone. The dashed
          line shows the peak $\rho-$ mass in vacuum inferred from
          $e^+e^-$ collisions (see inset for values of $M_R$ and
          $\Gamma_R$).  }
\label{fig:thermalshifts} 
\end{figure}
%
%
The peak position is determined by locating the value of $M$ at which
Eq.~(\ref{equ:noshifts}) exhibits a maximum. Setting the derivative of
Eq.~(\ref{equ:noshifts}) with respect to $M$ to zero, we solve for $x$
from
\beq{equ:derivative} 
\frac{4 x^2 (x^2-y^2)}{(x^2-y^2)^2+(y\,z)^2} +
\frac{x\, \sum_{j=1}^{\infty} {\rm K}_0(jx)} {\sum_{j=1}^{\infty}
\frac{1}{j} \, {\rm K}_1(jx)} -\frac{(2\ell+1) x^2}{x^2-4 y_\pi^2} = 1\,, 
\eeq
where we have introduced the abbreviations $x=M/T$, $y=M_R/T$,
$y_\pi=m_\pi/T$ and $z=\Gamma_R/T$ (for the $\rho-\pi\pi$~decay, $\ell
= 1$). For the temperatures chosen in Fig.~\ref{fig:thermalshifts},
Table 1 gives results for the peak positions $M_{\rm max}$, shifts
from the vacuum value $\Delta M = M_{\rm max}- M_R$, average masses
$\langle M \rangle$, and variances $\sigma_M = {\sqrt {\langle M^2
\rangle - \langle M \rangle^2 }}$, where the symbol $\langle \cdots
\rangle$ denotes averages taken over the invariant mass
distribution. Together with $M_{\rm max}$, the quantities $\langle M
\rangle$ and $\sigma_M$ provide a rough measure of the shape of the
distribution. As noted earlier, the shift of the peak mass is largest
at the smallest temperature, since in this case the steep Boltzmann
spectrum provides a strong bias towards low masses. This leads to the
intriguing possibility that a measured shift of the $\rho$-mass can be
used to delimit the freeze-out or decoupling temperature of the
disassembling hadronic fireball, provided medium-induced effects on
the $\rho$-spectral function and rescattering of the decay pions are
small or under full theoretical control.

The opposite pulls of the falling exponential (due to the Boltzmann
factor) and the rising power law (due to the phase space of decay
products and the pre-exponential factor in the non-degenerate density)
nearly cancel at $T \simeq 160$ MeV, leading to $\Delta M \simeq
0$. At the highest temperature considered ($T\simeq 175$ MeV is
thought to be the chemical freeze-out temperature at RHIC
\cite{BMMRS01}), the power law rise wins over the exponential fall
inasmuch as $\Delta M > 0$.  It is worthwhile remembering, however,
that at these temperatures, effects of interactions cannot be ignored,
and that the results shown in Fig.~\ref{fig:thermalshifts} and Table 1
refer to the effects of phase space alone.  Furthermore, as $T
\rightarrow T_c$, the basic structure of a resonance itself would be
considerably altered from its vacuum form due to effects of
deconfinement and/or chiral symmetry restoration.

%
\begin{center}
\vspace*{2mm}
 \begin{tabular}{|ccrcc|} \hline
   $\quad T \quad$    &  $\quad M_{\rm max} \quad$  &$\quad   \Delta M
\quad$
  & $\quad  \langle M \rangle \quad$ & $\quad \sigma_M \quad$\\  \hline \hline
   175   &   777.4         &      1.5     &    778.7      &  123.7
               \\
   150   &   774.9         &       $-1.0$     &    765.0      &  120.0   
            \\
   125   &   771.3         &       $-4.6$     &    746.4      &  118.9  
             \\ 
   100   &   765.3         &      $-10.6$     &    717.7      &  123.0 
              \\ \hline
 \end{tabular}
\end{center}
{\small Table 1: Basic characteristics of the $\rho-$ invariant mass
distributions in Fig.~~\ref{fig:thermalshifts}.  $T$ is the
temperature, $M_{\rm max}$ is the peak position, $\Delta M =
M_{\rm max}- M_R$ is the shift from the vacuum value, $\langle M
\rangle$ is the average mass, and $\sigma_M = {\sqrt {\langle M^2
\rangle - \langle M \rangle^2 }}$ is the variance of the distribution
(all values are in MeV).  The vacuum Breit-Wigner peak mass and width
are $M_R=M_\rho=775.9$ MeV and $\Gamma_R=\Gamma_\rho=147.9$ MeV,
respectively.
\label{tab:thermalshifts}}
\vspace*{5mm}

\suse{Analytic results to leading order}

%
To better understand the dependence of the shift on the decoupling
temperature, peak mass, and width of the resonance, we solve
Eq. (\ref{equ:derivative}) to linear order in the shift $\delta x
= x-y$, which is at best of order 10\% (see Table 1).  Keeping only
the first term in the sums in Eq. (\ref{equ:derivative}) (the Boltzmann
approximation is adequate around the peak mass), we find
\bea{equ:linshift}
\left(
\frac{8 \, y}{z^2} + \frac{24 \,y\, y_\pi^2}{(y^2-4y_\pi^2)^2}
+ y \left(\frac{{\rm K}_0}{{\rm K}_1}\right)^2
+ 2 \,    \frac{{\rm K}_0}{{\rm K}_1}
-y
\right) \delta x \nonumber \\
= \,
1 - y \,\frac{{\rm K}_0}{{\rm K}_1} + \frac{3y^2}{y^2-4y_\pi^2}\,,
\eea
where the Bessel functions are evaluated at $y=M_R/T$.  For 
$y^{-1} = T/M_R \ll 1$, we can simplify the above expression 
to obtain 
\beq{equ:linshift2}
\frac {\Delta M}{T}
\simeq 
\frac{-\left[ 1 - \left\{ \frac 32 + 3 
\left( 1 - \frac{4m_\pi^2}{M_R^2} \right)^{-1} 
\right\} \, \frac{T}{M_R} \right] }
     {8\, T^2/\Gamma^2 + T/M_R}\,.  
\eeq
In obtaining this result, use was made of the asymptotic expansion
${\rm K}_0(y)/{\rm K}_1(y) \simeq 1 - 1/(2y) + \cdots~$.  
Further linearization of
the fraction in Eq.~(\ref{equ:linshift2}) in terms of $T/M_R$ results
in
\beq{equ:linshift3}
\Delta M
\simeq - 
\frac{\Gamma^2}{8 T}
\left[1- \left\{ \frac 92 \left(1 + \frac 83 \frac {m_\pi^2}{M_R^2} \right) + 
\frac {\Gamma^2}{8T^2} \right\}  \frac {T}{M_R}   \right]\,. 
\eeq
Note that the second and third terms in Eq.~(\ref{equ:linshift2}) give
substantial contributions, which attests to the significance of the
power law behaviors of the available phase space (see
Eq.~(\ref{equ:momspace})) and the pre-exponential term in the density
of particles of mass $M$.  For example, the first term $\Delta M_0 = -
\Gamma^2/(8T)$ in Eq.~(\ref{equ:linshift3}) yields a shift of
\beq{equ:eg1}
\Delta M_0 =  -27.3~(-15.6)~\rm {MeV}~~~{\rm for}~~~T=100~(175)~\rm 
{MeV} \,,
\eeq
which is considerably larger in magnitude than --9.8(1.6) MeV given by
considering the other terms.  The corresponding shift from
Eq.~(\ref{equ:linshift3}) is --9.2(1.9) MeV. These shifts are to be
contrasted with the full numerical solutions from
Eq.~(\ref{equ:derivative}):
\beq{equ:eg2}
\Delta M = -10.6~(1.5~\rm {MeV}) \quad {\rm for} \quad
T=100~(175)~\rm {MeV} \,.
\eeq                               
Observe that Eq.~(\ref{equ:linshift2}) implies that $\Delta M = 0$ for
\beq{equ:zero}
T_z \simeq  \frac {M_R}{\frac 92 \left(1 + \frac {8m_\pi^2}{3M_R^2} \right)} 
\approx 160~{\rm MeV}  \,, 
\eeq
in excellent agreement with that obtained from
Eq.~(\ref{equ:derivative}).  A comparison of $M_{\rm max}$ versus
temperature from the different approximations is shown in
Fig.~\ref{fig:mversusT2b}. We have verified that effects of Bose
statistics are not very significant at these temperatures.
%

 \begin{figure}[t,b,p]
\epsfig{file=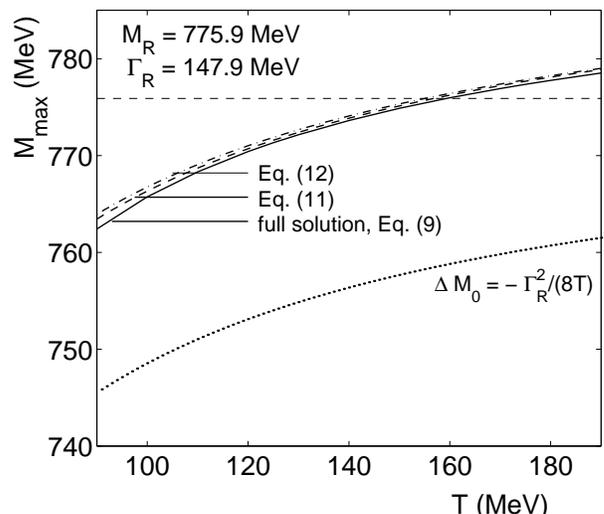,width=8cm}
 \caption{Peak positions of the invariant mass distribution functions
          in Fig.~1 versus temperature. The solid curve refers to the
          exact numerical results from Eq.~(\ref{equ:derivative}),
          while the dashed and dash-dotted curves are for the
          approximations in Eqs. (\ref{equ:linshift2}) and
          (\ref{equ:linshift3}), respectively.  The dotted curve
          refers to the zeroth order (in $T/M_R$) result $\Delta
          M_0=-\Gamma^2/ 8T$.  }
\label{fig:mversusT2b} 
\end{figure}
%

\suse{ Shifts induced by in-medium interactions plus thermal
and decay product phase spaces}  

From Eqs.~(\ref{equ:linshift2}) and (\ref{equ:linshift3}), it is
evident that the width of a resonance strongly influences the
magnitude of the peak mass shift.  Thus, a broad resonance such as the
$f_0(600)$ will exhibit a larger shift than sharp resonances such as
the $K^*$ or $\phi$.  In-medium interactions with the other
constituents of the heat bath, namely mesons such as $\pi, K, ..,
\omega, \phi, a_1, ...$ and baryons such as $ \Delta, N^* ...$, will
likely modify the $\rho-$spectral function from its vacuum form.  The
observed distributions are, however, convolutions of such intrinsic
changes (pressure broadening and interaction-induced mass shift) with
the thermal phase space and the phase space for the corresponding
decay products. These intrinsic changes are given by the self-energy
$\Pi$ (see Eq.~(\ref{equ:self-energy})), with ${\rm Re}~\Pi$ giving
the intrinsic mass shift and ${\rm Im}~\Pi$ giving the in-medium
width.

Before turning to detailed model calculations of $\Pi$, it is useful
to explore the combined effects of schematic intrinsic changes (such
as a broadened, but unshifted resonance or a shifted, but not
broadened resonance; the realistic case would include a combination of
both) and the thermal and decay product phase space functions.  In
Fig. \ref{fig:gammashifts}, we show the invariant mass distribution of
the $\rho$-meson with a fixed peak mass of 775.9 MeV at a temperature
of 100 MeV for different widths ($\Gamma_R = 100$ to 200 MeV in steps of
25 MeV). It is evident that an intrinsic broadening of the resonance
induces a considerable shift in the peak mass; the larger the broadening,
the larger is the shift.

%
 \begin{figure}[t,b,p]
 \epsfig{file=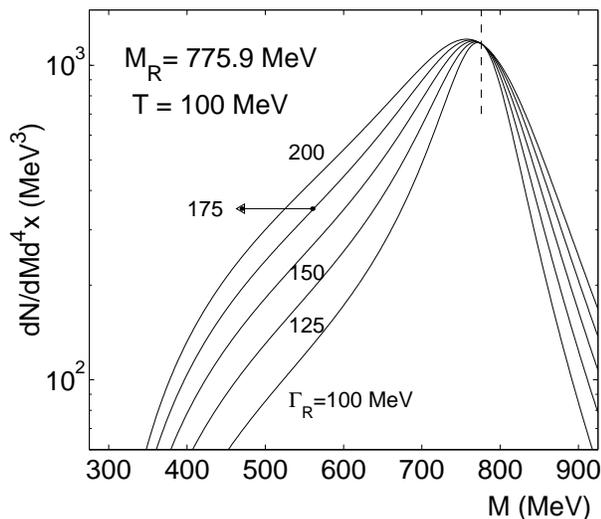,width=8cm}
 \caption{Invariant mass distributions of the $\rho$ meson in a
          hadronic heat bath (at a temperature of 100 MeV) for varying
          intrinsic widths.  The dashed line indicates the unshifted
          rho-mass $M_R=M_\rho=775.9$~MeV.  } 
\label{fig:gammashifts}
\end{figure}
%

It is noteworthy that for low temperatures and large intrinsic widths,
the shape of the invariant distribution is considerably distorted from
its symmetric form in vacuum.  Consequently, the definition of the
``width'' as normally used for symmetric shapes loses its utility as
shown by the standard deviations $\sigma_M$ in Table 2.  The
locations of the maxima and the mass-shifts are also listed in Table 2
for the various curves shown in Fig.~3.

%
\begin{center}
\vspace*{2mm}
 \begin{tabular}{|ccrcc|} \hline
 $\quad \Gamma_R \quad$&  $\quad M_{\rm max} \quad$  &  $\quad
\Delta M 
\quad$ &  $\quad \langle M \rangle \quad$   & $\quad \sigma_M \quad$      \\  
\hline \hline
    100           &   771.3              &     $-4.6$            &  732.4      
     &    107.3                  \\
    125           &   768.7              &     $-7.2$            &  724.3     
      &    116.2                  \\
    150           &   765.4              &     $-10.5$           &  717.2     
      &    123.6                  \\
    175           &   761.6              &     $-14.3$           &  710.7     
      &    129.8                  \\
    200           &   757.2              &     $-18.7$           &  705.0     
      &    135.1                  \\ \hline
 \end{tabular}
\end{center}
{\small Table 2: Shifts (from Eq. (\ref{equ:derivative})) of the peak
mass induced by varying intrinsic widths, $\Gamma_R$, of the
resonance. The Breit-Wigner peak mass is at $M_R=M_\rho=775.9$~MeV.
Also shown are the average mass and the standard deviation of the
distribution.  All values are in MeV. }
\vspace*{5mm}
%

We turn now to the case in which intrinsic changes in the peak mass
alone are considered. For simplicity, and because the dependence on
the momentum of the resonance is small for low temperatures and baryon
number densities at freeze-out (see Fig.~8 of Ref.~\cite{EBEK02}), we
use constant peak mass shifts for illustrative purposes.  The results
in Fig. \ref{fig:massshifts} and Table 3 confirm our expectations from
Eqs.~(\ref{equ:linshift2}) and (\ref{equ:linshift3}) that the mass
shift is largely independent of the peak mass. For the case
considered, $T=100$ MeV, the mass shift relative to the peak mass
varies from 1.4\% for $M_R=836$~MeV to 1.1\% for $M_R=716$~MeV.  Thus,
relative to the assumed peak mass, shifts induced by intrinsic mass
shifts are smaller than those caused by changes in the width.

%
 \begin{figure}[t,b,p]
 \epsfig{file=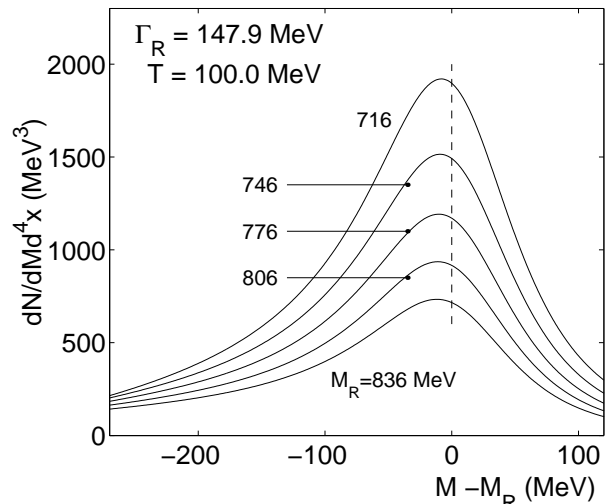,width=8cm}
 \caption{Invariant mass distributions of the $\rho-$ meson with a
          width $\Gamma_R=\Gamma_\rho=147.9$~MeV 
          and assumed intrinsic peak masses
          that vary from 716 to 836 MeV.  For each curve, $M_R$ is
          subtracted from the horizontal scale.  }
 \label{fig:massshifts}
 \end{figure}
%

%
\begin{center}
\vspace*{2mm}
 \begin{tabular}{|ccrcc|} \hline
$\quad M_R \quad$ &  $\quad M_{\rm max} \quad$ &  $\quad \Delta M
\quad$ & 
$\quad \langle M \rangle \quad $ & $\quad \sigma_M \quad$ \\  \hline \hline
   716         &    707.8           &     $-8.2$           &    674.7
& 
    111.3        \\
   746         &    736.7           &     $-9.3$           &    696.4
&
     116.9        \\
   776         &    765.8           &    $-10.2$          &    717.8
& 
    123.0        \\
   806         &    795.0           &    $-11.0$           &    738.7
& 
    129.7        \\
   836         &    824.2           &    $-11.8$           &    759.1
& 
    137.0        \\ \hline
 \end{tabular}
\end{center}
{\small Table 3: Peak mass $M_{{\rm max}}$ and shift 
$\Delta M  = M_{\rm max} - M_R$ (from Eq. (\ref{equ:derivative}))
versus the Breit-Wigner maximum $M_R$ for a fixed 
width $\Gamma_R = \Gamma_\rho =
147.9$~MeV (all numbers are in MeV).  Also shown is the 
the standard deviation $\sigma_M$ of the invariant distribution.  }
\vspace*{5mm}
%

The results shown in Fig.~\ref{fig:massshifts} and Table 3 refer to
peak mass shifts caused by the phase space functions on top of the
assumed intrinsic mass shifts in the range $-60 \leq \Delta M \leq 60$
MeV~{\footnote{Upward intrinsic mass shifts are also considered to
encompass the expectations of Ref.~\cite{Pisarski95}.}}.  Compared to
the vacuum peak mass, the total shift is given by
\beq{equ:totshift}
\Delta M_{tot} = (M_R-775.9) + \Delta M \,.
\eeq 
Thus, for example, 
\bea{equ:toteg}
\Delta M_{tot} &=& -68.1~{\rm MeV}\quad {\rm for}~ 
M_R=716~{\rm MeV} \,, \nonumber \\
&=& + 48.3~{\rm MeV}\quad {\rm for}~
M_R=836~{\rm MeV} \,.
\eea
It is evident that sizeable intrinsic mass shifts are required to
generate sizeable peak mass shifts in the invariant mass
distributions.

\se{Contributions to the $\pi\pi$ final state from the 
$\omega(782)$ and  $f_0(600)$ mesons}

In the peak region of the $\rho-$invariant mass spectrum,
contributions to the $\pi\pi$ final state include those that arise
from the $f_0(600)$ (hereafter $\sigma$) and $\omega$ mesons that are
also likely modified by medium effects.  

\suse{$\omega \rightarrow \pi^+\pi^-$ decays} 

The decay of the $\omega$ is distinctive, chiefly because of its
narrow total width $\Gamma = 8.44$ MeV in vacuum.  Although the
dominant decay channel is the $\omega \rightarrow \pi^+\pi^-\pi^0$
decay ($\Gamma_i/\Gamma = 89.1 \pm 0.7 \%$), a small fraction
($\Gamma_i/\Gamma = 1.70 \pm 0.28 \%)$ of the decays occur via the
$\omega \rightarrow \pi^+\pi^-$ channel.  For the latter case,
Eqs.~(\ref{equ:derivative}) through (\ref{equ:linshift3}) show that
the influence of the phase space functions on the peak mass shift of
the $\omega$ is negligibly small.  Unless large intrinsic shifts are
caused by the medium, the shift in the $\omega-$peak mass in the
invariant distribution of its $\pi^+\pi^-$ final state is likely to be
very small.  Calculations to date~\cite{EBEK02,KKW97} have found only
modest changes in ${\rm Re}~\Pi_\omega$ and ${\rm Im}~\Pi_\omega$ at
the freeze-out temperatures of (100--125) MeV.

\suse{$\sigma \rightarrow \pi^+\pi^-$ decays} 

Pinning down the contributions from the $\sigma$ decays is more
difficult, especially since its vacuum width is rather large
$(600-1000~{\rm MeV})$ and its peak position $(400-1200~{\rm MeV})$ in
vacuum is somewhat uncertain~\cite{PDG02}. To encompass the possible
variations, we first present results of two illustrative calculations; 
one in which we take $M_R(\sigma) = 800$ MeV and $\Gamma_R(\sigma) =
800$ MeV (this case refers to the average peak mass and the average
width, respectively), and the other in which we employ $M_R(\sigma) =
600$ MeV and $\Gamma_R(\sigma) = 400$ MeV (these values represent
averages from the $D-$ and $\tau-$ decays~\cite{PDG02}).  The vacuum
decay width for the $\sigma \rightarrow \pi\pi$ is 
\bea{equ:smomspace}
\Gamma_{R \rightarrow f}^{\rm vac}(\sigma) &=& 
\Gamma_R(\sigma) \left(\frac {k}{k_0}\right)^{2\ell +1} \nonumber \\ 
&=& \Gamma_R(\sigma)
\left(\frac{M^2 - 4 \, m_\pi^2 } {M_R^2(\sigma)-4\,m_\pi^2}\right)^{1/2}\,, 
\eea 
as, in this case, the orbital angular momentum $\ell = 0$.  Utilizing
this in Eq.~(\ref{equ:noshifts}) with spin $J=0$, results of invariant mass
distributions from the $\sigma\rightarrow \pi\pi$ decay are shown by the
dash-dotted lines in Fig.~5 for temperatures expected near freeze-out.
As is clear from the solid curves in this figure, which
show the summed contributions from the $\rho \rightarrow \pi\pi$
(dashed curves) and $\sigma \rightarrow \pi\pi$ decays, the maximum
near the peak-region of the $\rho-$ mass spectrum is barely influenced
by the small and exponentially falling contributions from the $\sigma
\rightarrow \pi\pi$ decays.  The contributions from the $\sigma
\rightarrow \pi\pi$ decays dominate, however, for masses well below
the $\rho-$ peak.

By setting $\ell=0$ in Eq.~(\ref{equ:derivative}), one can
straightforwardly obtain a very rough estimate of the peak mass shift
in the invariant mass distribution from the $\sigma \rightarrow
\pi\pi$ decay. The result to linear order in $T/M_R$ is
\beq{equ:Sshift}
\frac {\Delta M}{T}
= 
\frac{-\left[ 1 - \left\{ \frac 32 + 
\left( 1 - \frac{4m_\pi^2}{M_R^2} \right)^{-1} 
\right\} \, \frac{T}{M_R} \right] }
     {8\, T^2/\Gamma^2 + T/M_R}\,.  
\eeq
This result, however, provides a poor approximation to the numerical
results from Eq.~(\ref{equ:derivative}) chiefly because the shifts are
large.

 \begin{figure}[t,b,p]
 \epsfig{file=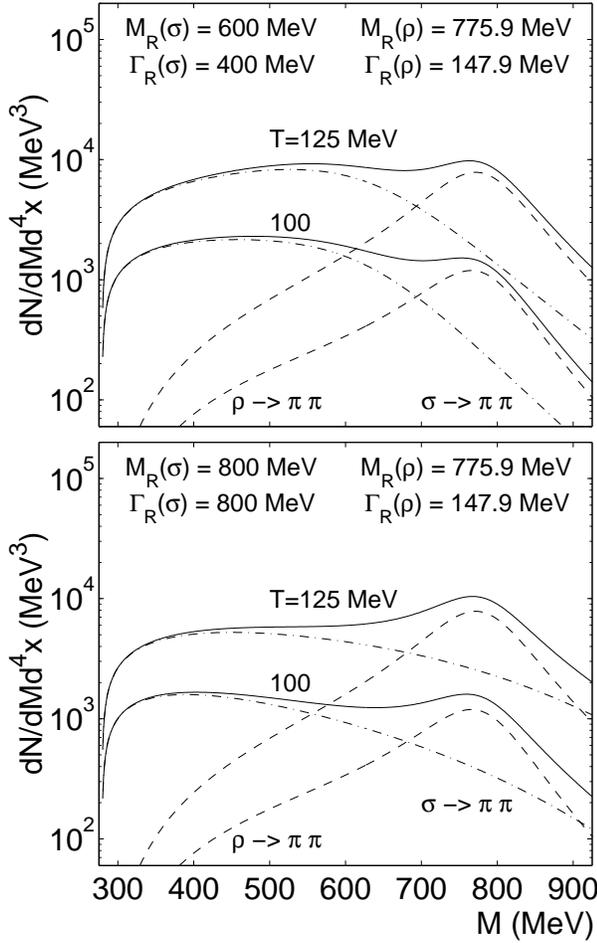,width=8cm}
 \caption{Invariant mass distributions from the decays of the $\rho$
	(dashed curves) and $\sigma$ (dash-dotted curves) mesons.  The
	solid curves show the sum of the contributions. The
	Breit-Wigner maximum masses and widths are shown in the
	insets.  }
\label{fig:rho_sig}
\end{figure}
%

Table 4 lists the peak masses $M_{{\rm max}}$ and shifts $\Delta M$
calculated numerically from Eq. (\ref{equ:derivative}) with $\ell=0$
for the cases $M_R(\Gamma_R)$ 600 (400) and 800 (800) MeV,
respectively.  Notice that the shifts in the invariant mass
distributions from the $\sigma-$ decay in medium depend strongly on
the parameters used to describe the shape of the $\sigma-$ resonance
in vacuum.

%
\begin{center}
\vspace*{2mm}
 \begin{tabular}{|c|ccr|} \hline
 &   $\quad T \quad$ & $\quad M_{\rm max} \quad$ &  $\quad \Delta M
\quad$  \\  \hline \hline
$M_R$ = 600       & 125 & 527.2 & $-72.8$ \\       \cline{2-4}
$\Gamma_R$ = 400  & 100 & 458.5 & $-141.5$      \\ \hline
$M_R$ = 800       & 125 & 450.5 & $-349.5$ \\      \cline{2-4}
$\Gamma_R$ = 800  & 100 & 391.0 & $-409.0$     \\ \hline \end{tabular}
\end{center}
{\small Table 4: Peak mass $M_{{\rm max}}$ and shift 
$\Delta M  = M_{\rm max} - M_R$ (from Eq. (\ref{equ:derivative}))
versus temperature $T$ for 
the indicated values of the Breit-Wigner maximum $M_R$ and  
width $\Gamma_R$ (all numbers are in MeV). } 
\vspace*{5mm}
%

\suse{In-medium shifts of the $\sigma-$meson} 

The intrinsic mass shift of the fragile $\sigma-$meson is intimately
related with chiral symmetry restoration (cf. \cite{HK94} and
references therein).  With increasing temperature (or baryon density),
$M_R(\sigma)$ is expected to steadily decrease and approach the pion
mass $m_\pi$ as chiral symmetry is fully restored at $T_c \simeq \ 170 \pm
10$ MeV (the precise number is still a matter of intense debate!).
Being a Goldstone boson, $m_\pi$ remains nearly constant at its vacuum
value even as $T_c$ is approached.  Thus, at the freeze-out
temperatures $T \sim 100-125$ MeV and rather small baryon densities
under consideration here, modest downward shifts of the $\sigma-$ peak
mass from its vacuum value are to be expected.  For definiteness, we
choose an optimistic intrinsic mass shift of $100$ MeV relative to its
vacuum value of 600 MeV based on the model calculations reported in
\cite{HK94}. At first, we take $\Gamma_R(\sigma)=400$ MeV. In this
case, the shifts relative to $M_R=500$ MeV are $\Delta M = -73 (-37)$
MeV for $T=100(125)$ MeV, respectively.  These shifts, relative to the
vacuum peak mass of 600 MeV, amount to the total shifts
\bea{equ:stoteg1}
\Delta M_{tot} &=& -173~{\rm MeV}\quad {\rm for}~ 
T=100~{\rm MeV} \,, \nonumber \\
&=& - 137~{\rm MeV}\quad {\rm for}~
T=125~{\rm MeV} \,.
\eea
In addition to its mass, the $\sigma-$decay width $\Gamma_R(\sigma)$
also decreases with increasing temperature.  For illustration, we
employ $\Gamma_R(\sigma)=300$ MeV. This results in shifts $\Delta M
= -46.9 (-23.2)$ MeV for $T=100(125)$ MeV. The corresponding total
shifts are
\bea{equ:stoteg2}
\Delta M_{tot} &=& -146.9~{\rm MeV}\quad {\rm for}~ 
T=100~{\rm MeV} \,, \nonumber \\
&=& - 123.2~{\rm MeV}\quad {\rm for}~
T=125~{\rm MeV} \,.
\eea

In all cases considered, the $\sigma \rightarrow \pi\pi$ contributions
at the peak $\rho-$mass are given predominantly by the exponentially
falling Boltzmann factors and provide small corrections to those from
$\rho \rightarrow \pi\pi$ decays. The left-wings of the
$\rho-$distributions are, however, destroyed insofar as the summed
invariant distributions are relatively flat over a wide range of
invariant masses $M \leq 4.5m_\pi$. Note, however, that the $K_0^s(498)
\rightarrow \pi\pi$ decays that essentially occur in vacuum exhibit a
prominent peak at $\sim 500$ MeV and that the $\sigma \rightarrow
\pi\pi$ decays constitute a background in the vicinity of this peak.

\se{Shifts of baryon resonances}

Most of the secondary particles produced in RHIC collisions are
mesons.  However, many baryons and baryon resonances including their
anti-particles are also produced.  Invariant distributions of hadronic
decay products from baryon resonances are best exemplified by
considering the decay of the $\Delta(1232)$-resonance. In-medium
properties of the $\Delta$ have been studied earlier through
photo-absorption and hadronic reactions on
nuclei~\cite{BW75,OTW82,EW88}, but has received relatively little
attention at temperatures characteristic of RHIC collisions.

\suse{Effects of thermal motion and the phase space of unequal mass
decay products} 

As in the case of $\rho \rightarrow \pi\pi$ decay, the $\Delta
\rightarrow n \pi$ decay, where $n$ is a nucleon, occurs with orbital
angular momentum $\ell = 1$.  Thus, the invariant mass distribution can be
calculated following the procedure outlined in Sec. II, but with
\beq{equ:DGamma}
\Gamma_{R \rightarrow f}^{\rm vac} = \Gamma_R
\left( \frac {k}{k_0}\right)^{2\ell +1} \,,
\eeq
where $k$ is the c.m. momentum of either the pion or nucleon, {\it i.e.},  
\beq{equ:cmk}  
k = \frac 12    
\left(\frac {(M^2 -  M_+^2)  (M^2 -  M_-^2)} {M^2}\right) ^{1/2} \,, 
\eeq 
where $M_\pm = M_n \pm m_\pi$ and $k_0$ is the c.m. momentum with
$M=M_R$. (Near the peak region, the functional form in
Eq.~(\ref{equ:DGamma}) matches that advocated in
Ref.~\cite{EW88} in which a more complete discussion  of 
$\Gamma_{R\rightarrow f}^{{\rm vac}}(\Delta)$ can be found.)
In this case, the maximum in the invariant mass distribution is
obtained by solving for $x$ from
\bea{equ:Dderiv} 
\frac{4 x^2 (x^2-y^2)}{(x^2-y^2)^2+(y\,z)^2} +
\frac{x\, \sum_{j=1}^{\infty} {\rm K}_0(jx)} {\sum_{j=1}^{\infty}
\frac{1}{j} \, {\rm K}_1(jx)}  \nonumber \\ 
~~~~~- \frac {3(x^4-y_+^2 y_-^2)} {(x^2 - y_+^2) (x^2 - y_-^2)} = 1\,, 
\eea
where $x=M/T,~y=M_R/T,~y_{\pm}=M_{\pm}/T$, and $z=\Gamma_R/T$. Since
for masses near the peak region, $M/T \gg 1$, Eq.~(\ref{equ:Dderiv})
can be solved for the linear shift $\delta x = x-y$ as in Sec.  II.B
by considering only the $j=1$ terms in the second term (the Boltzmann
limit).  To linear order in $T/M_R$, the shift in the peak mass is
given by
\beq{equ:Dshift}
\Delta M \simeq - \frac {\Gamma^2}{8T} 
\left[1 - \left(\frac 32 + 3 Q + 
\frac {\Gamma^2}{8T^2} \right) \frac {T}{M_R}  \right] \,, 
\eeq
where 
\beq{equ:CQ}
Q = \frac {M_R^4- (M_+M_-)^2}{(M_R^2 - M_+^2) (M_R^2 - M_-^2) } 
\eeq
Note that Eq.~(\ref{equ:Dshift}) reduces to Eq.~(\ref{equ:linshift3})
in the limit of identical final state particles.
The extent to which the linear shift in  Eq. (\ref{equ:Dshift})
reproduces the numerical results from  Eq. (\ref{equ:Dderiv}) is
shown in Table 5. 

%
\begin{center}
\vspace*{2mm}
 \begin{tabular}{|c|cc|} \hline
$\quad T \quad$ & $\quad  \Delta M ( {\rm Eq}.~(\ref{equ:Dderiv}))
\quad$  &  $\quad \Delta M ({\rm Eq}.~(\ref{equ:Dshift})) \quad$  \\  \hline \hline
175 & 12.6 & 13.8 \\
150 & 11.0 & 12.0      \\ 
125 & 8.7 &   9.6 \\
100 & 5.4 &   6.1     \\ \hline \end{tabular}
\end{center}
{\small Table 5: Comparison of peak mass shifts $\Delta M = M_{\rm
max} - M_R$ from Eq. (\ref{equ:Dderiv})) and Eq. (\ref{equ:Dshift}))
versus temperature $T$ for the $\Delta^0 \rightarrow p\pi^-$ decay.
The Breit-Wigner maximum $M_R=1232$ MeV and $\Gamma_R=120$ MeV. The
shifts are in MeV. }
\vspace*{5mm}
%

The temperature at which the shift $\Delta M=0$ is given by
\beq{equ:Dzero}
T_z \simeq \frac {M_R}{\frac 32 + 3Q} \approx 74.5~{\rm MeV}.
\eeq
This temperature, being considerably lower than the freeze-out
temperature $100 < T_f/{\rm MeV} < 125$, implies that an observed
downward shift from the vacuum peak value of 1232 MeV from
$\Delta-$decay distinctively signals an intrinsic downward shift due
to in-medium effects~\footnote{The precise value of $T_z$ depends on
the peak mass and width of the $\Delta-$resonance.  Intrinsic shifts of these
quantities considered in Sec. IV.B result in values of $T_z \simeq
75\pm 25$ MeV. Thus, the main conclusion that $T_z \leq T_f$ remains
valid in the presence of medium modifications of the width and peak
mass of the $\Delta-$resonance.}. By the same token, an observed
upward shift must be attributed to the predominance of the phase space
function associated with the decay products, although the left-wings
of higher mass $N^*$ resonances can also contribute in determining the
peak position. We turn now to effects of in-medium mass and width
modifications on the peak mass shift.

\suse{In-medium shifts of the $\Delta-$ isobar} 

In photonuclear reactions, cross-sections per nucleon for nuclei
ranging from $^9$Be to $^{238}$U when compared to the average single
nucleon total photoabsorption cross-section clearly show that the
$\Delta-$resonance appears at an energy close to that of the free
$\Delta(1232)$, but spread over a wider energy
range~\cite{Ahrens85}.  In $\pi-$nucleus reactions, however, in
addition to a widened width, the resonance energy is also shifted
downward by about 20--30 MeV. An illuminating discussion of the
various effects contributing to in-medium shifts at zero
temperature can be found in~\cite{EW88}. The total width arises from
a combination of competing effects:
\beq{equ:MDwidth} \Gamma (\Delta) = \Gamma_{\Delta}^{\rm free} +
\Gamma^{\rm el} + \Gamma^{\rm abs} + \delta\Gamma_\Delta^{\rm Pauli} \,.
\eeq
The dominant contribution arises from the so-called elastic broadening
of the width, $\Gamma^{\rm el}$, given by the imaginary part of the
matrix element $\langle (\Delta h)_\beta | V_\pi | (\Delta h)_\alpha
\rangle$, where $h$ stands for a nucleon hole and $V_\pi$ denotes the
pion-mediated interaction between baryons 1 and 2. Coupling to
absorptive channels (typically one-quarter to one-third of the total
cross-section in the resonance region comes from this source)
corresponds to the coupling of $\Delta-$hole states to multiple
nucleon-hole (nNnh) states and increases the width. In contrast, the
width is reduced by the Pauli exclusion principle, since in the decay
$(\Delta h) \rightarrow \pi (Nh) $, occupied nucleon states are
inaccessible. The net effect, however, is a net increase in the width
of $(\Delta h)$ states. For low baryon densities and for temperatures
of interest here, however, the dominant contributions arise from
$\Gamma^{\rm el}$ and $\Gamma^{\rm abs}$, since Pauli effects are
small for non-degenerate baryons. Additional modifications from
coupling to higher-lying $N^*$-resonances remain largely unexplored.

The width modifications are accompanied by dispersive energy shifts
(for nuclei, these amount to a few tens of MeV). Such shifts are due
to the combined effects of $\Delta-$binding in matter, dispersive
shifts related to absorption, short-range $\Delta-$hole correlations,
coupling to higher-lying $N^*$-resonances, {\em etc}.  To date, the
individual contributions from these various sources have not been
resolved in experiments.

%
\begin{center}
\vspace*{2mm}
 \begin{tabular}{|c|ccc|} \hline
 &   $\quad T \quad$ & $\quad \Delta M \quad$ &  $\quad \Delta M_{\rm tot}
\quad$  \\  \hline \hline
                 & 125 & 14.2 & $-25.9$ \\
\raisebox{1.5ex}[-1.5ex]{$M_R$ = 1192 }
                 & 100 & 10.9 & $-29.1$      \\ \hline
                 & 125 & 11.1 & $- 8.9$  \\ 
\raisebox{1.5ex}[-1.5ex]{$M_R$ = 1212 }
                 & 100 &  7.9 & $-12.2$      \\ \hline
                 & 125 &  6.8 & $ 26.8$  \\  
\raisebox{1.5ex}[-1.5ex]{$M_R$ = 1252 }
                 & 100 &  3.5 & $ 23.5$      \\ \hline
                 & 125 &  5.1 & $ 45.1$  \\  
\raisebox{1.5ex}[-1.5ex]{$M_R$ = 1272 }
                 & 100 &  1.8 & $ 41.8$      \\ \hline
\end{tabular}
\end{center}
{\small Table 6: Peak mass shifts $\Delta M$ from
Eq.~(\ref{equ:Dderiv}) caused by in-medium Breit-Wigner peak values of
$M_R$ and the vacuum value of $\Gamma_R=\Gamma_\Delta=120$ MeV.  
The values of $T$
bracket the likely temperature at freeze-out.  $\Delta M_{\rm tot} =
(M_R - 1232\, {\rm MeV}) + \Delta M$ refers to the total shift from the vacuum
peak value.  All numbers are in MeV.}
\vspace*{5mm}
%

%
\begin{center}
\vspace*{2mm}
 \begin{tabular}{|c|cc|} \hline
 &   $\quad T \quad$ & $\quad \Delta M \quad$   \\  \hline \hline
                 & 125 &  6.3        \\ 
\raisebox{1.5ex}[-1.5ex]{$\Gamma_R=100$ }
                 & 100 &  3.9   \\ \hline  
                 & 125 &  11.5    \\  
\raisebox{1.5ex}[-1.5ex]{$\Gamma_R=140$ }
                 & 100 &  7.1 \\ \hline  
\end{tabular}
\end{center}
{\small Table 7: Peak mass shifts $\Delta M$ from
Eq.~(\ref{equ:Dderiv}) due to in-medium Breit-Wigner widths $\Gamma_R$ and
the vacuum peak value of $M_R=M_\Delta=1232$ MeV.  The values of $T$ bracket
the likely temperature at freeze-out.  All numbers are in MeV.}
\vspace*{5mm}

Tables 6 and 7 show results of shifts $\Delta M$ and 
$\Delta M_{\rm tot} = (M_R - 1232\,{\rm MeV}) + \Delta M$ 
calculated from Eq.~(\ref{equ:Dderiv})
with representative in-medium shifts.  These results show how
medium-induced modifications are reflected in the invariant
distribution of $p\pi$ final states from $\Delta-$ decays
alone. Viewed together with the results in Table 5, the results in
Tables 6 and 7 imply that a downward shift in the peak value of the
$\Delta-$resonance requires a substantial dispersive energy shift in
the medium. Building upon the known $\pi N\Delta$ physics, an in-depth
finite temperature analysis that includes coupling to higher $N^*$
resonances is necessary.

\se{ Discussion}
  
\suse{In-medium spectral functions}

In the context of lepton-pair ({\it e.g.}, $e^+e^-$) and photon
production from
the dense stages of heavy-ion collisions, considerable amount of work
has been done on the in-medium spectral functions of vector mesons
(see, {\it e.g.}, Refs.~\cite{Gale02,RW00} for exhaustive references).
The viewpoints that have emerged can be roughly classified into two
broad categories:

\noindent (1) The masses of the vector mesons, particularly those of
the $\rho$ and $\omega$, decrease substantially with increasing
temperature $T$ and net baryon density $n_B$. This view, espoused in
Ref.~\cite{BR91} and shared to varying degrees in other
works~\cite{HK94,HY02}, is based essentially on the decreasing
behavior of the $\langle \bar qq \rangle$ condensate which in turn is
related to the vector meson masses as a function of $T$ and $n_B$.
For additional dependences on strong interaction couplings, see
Refs.~\cite{BR91,HK94,HY02}.

\noindent (2) The peaks of the spectral densities are little shifted
from their vacuum positions, but the widths are considerably increased
due to medium effects that include collisional
broadening~\cite{Gale02,RW00,EBEK02}. This view stems from many-body
calculations that consider the many hadron states (for an extensive
list, see Sec. II of Ref.~ \cite{EBEK02}, and particularly, Table 1
therein) to which the vector mesons couple in a heat bath.  By relating
the self-energy to the real and imaginary parts of the forward
scattering amplitude, Ref.~ \cite{EBEK02} exploits experimental data
where possible to infer the peak positions and widths in medium.

In both scenarios, the largest medium-induced effects, whether it is a
mass shift or a widened width, occur for $T \rightarrow T_c \simeq 170
\pm 10$ MeV and for $n_B \gg n_0$, where $n_0 \simeq 0.16~{\rm
fm}^{-3}$ is the nuclear equilibrium density. However, 
near freeze-out $T \simeq (100-125)~{\rm MeV}
\ll T_c$ and $n_B \ll n_0$.  Since our focus here is on the properties
of resonances as reflected by their hadronic decay products at the
dilute stages of freeze-out, medium-induced effects are expected to be
considerably smaller than those for lepton-pair and photon production which
occurs dominantly at the dense stages of the collision.

In computing shifts of peaks in the invariant mass distributions in
this work, we have utilized the basic trends found in the intrinsic
shifts of both the peak masses and widths in both of the above
approaches. Specifically, the effects of widened widths and both
downward and upward intrinsic mass shifts on the invariant mass
distributions were investigated.  For temperatures at freeze-out and
negligibly small baryon densities under consideration, the momentum
dependences of the intrinsic widths are rather weak~\cite{EBEK02}
being significant only for momenta $q \geq q_{av} \simeq {\sqrt
{3MT}}$.  These weak momentum dependences are washed out upon
convolution with thermal and decay product phase space functions.
Thus, our discussion has been carried out with momentum-independent
in-medium widths.  For the decay widths, however, the
appropriate orbital angular momentum dependences were considered.

\suse {Flow--related matters}

Hadrons of various species ($\pi, K, N, ~etc., $) are known to exhibit
strong collective expansions (large longitudinal and transverse flows
have become a part of RHIC life; see \cite{flowreviews} and references
therein) as a result of pressure gradients established at a relatively
early stage of the collision \cite{rapidtherm}.  Prime examples of
collective flow include the ``blue-shift'' or flattening of the
observed transverse momentum spectra \cite{SSH93}, and the azimuthal
angular anisotropy or elliptic flow \cite{Ollitrault92}, of the
various particles.  Invariant mass spectra, however, do not depend on
Lorentz boosts and are thus unaffected by collective flow.
Contributions from the various phase space cells can be determined in their
rest frames and subsequently summed to give the total
distribution. This underscores the importance of spectral function
studies that are potentially the only means to delineate the intrinsic
properties of resonances at both the dense and freeze-out stages of
high-energy heavy-ion collisions.

\suse{Relation to other works}

Based on the nonrelativstic expression for the Breit-Wigner peak
\beq{equ:NBW}
\rho(M) \propto \frac {M\Gamma}{(M-M_R)^2 + \Gamma^2/4} \,,
\eeq  
Ref.~\cite{BBFSB91} found a downward energy shift of the maximum of the
energy spectrum of the $\rho-$mesons to be 
\bea{equ:bform}
\Delta M &\simeq& - \frac {\Gamma^2}{8T} 
\left( 1 - \frac 32 \frac {T}{M_R}   \right) \nonumber \\
&\approx& -17.5~{\rm MeV} \quad {\rm for} \quad T=120~{\rm MeV} \,.
\eea
This shift differs substantially from the $-5.4$ MeV shift at the same
temperature obtained in this work.  It is easy to verify that shifts
to linear order obtained from the use of nonrelativistic and
relativistic Breit-Wigner forms both lead to the same result in
Eq.~(\ref{equ:linshift}). We therefore surmise that the substantial
positive shift from the third term in the numerator of
Eq.~(\ref{equ:linshift2}),
\bea{equ:sps}
\left(\Delta M\right)_{\rm phase~space} &\simeq& 
\frac {\Gamma^2}{8T} \cdot 3 \left(1 - \frac {4m_\pi^2}{M_R^2} \right)^{-1}    
\cdot \frac {T}{M_R} 
\nonumber \\
&\approx& 12.2~{\rm MeV} \quad {\rm for} \quad T=120~{\rm MeV} \,.
\eea
which arises from the phase space of the decay products, was possibly
not fully considered in estimating the shift in Ref.~\cite{BBFSB91}.

Our work here partly overlaps that in Ref.~\cite{SB02}.  Complementary
information, including mass shifts through the Brown-Rho scenario, the
role of $t-$channel resonances in inducing mass shifts of the
$\rho-$meson, some aspects of kinetic freeze-out, effects of the heat
bath {\em etc.,} can be found in Ref.~\cite{SB02}.  However, the
influence of the angular-momentum-dependent decay product phase space,
crucial for comparison with experimental invariant mass distributions,
was not considered in~\cite{SB02}.  Our work, in addition to
bracketing the effects of different scenarios concerning
medium-induced effects, highlights the importance of the competition
between intrinsic shifts and phase space functions related with
thermal effects and decay products.  The large differences in the peak
mass shifts in invariant mass distributions between our work and those
of Refs.~\cite{BBFSB91,SB02} stem primarily from the fact that the
decay product phase space functions were not considered in the latter
works.

The results in Secs. II through IV clearly show that qualitative
differences exist in the influence of the $M-$dependence on the $\ell
= 1$ decays of $\rho \rightarrow \pi\pi$ and $\Delta \rightarrow
p\pi$, insofar as the decay product phase space functions, in
conjunction with thermal phase space and expected intrinsic shifts,
produce qualitatively different results.  In $\rho \rightarrow \pi\pi$
decay, the net result is a downward shift from the vacuum value of the
$\rho-$peak mass at freeze-out temperatures.  In $\Delta \rightarrow
p\pi$ decay, however, larger than vacuum peak masses are favored
unless large modifications in the intrinsic mass and width tip the
balance.  Our work shares the view with that of Ref.~\cite{SB02} that
substantial in-medium collisional and dispersive energy shifts are
necessary to produce the substantial mass shift in the reported
$\rho-$meson invariant mass spectrum \cite{Fachini02}.

\se{Summary and principal findings }

A study of the peak positions of resonances and their shapes is a
time-honored spectroscopic tool that is beginning to be used to probe
the freeze-out conditions at RHIC collisions and to establish the
properties of resonances in a hot and dense medium.  The recently
reported 60--70 MeV downward shift in the peak of the invariant mass
distribution of pions from the decay of $\rho(776)$ mesons is
particularly interesting in this regard, since the decay products
originate from the dilute freeze-out stage of the collision.  In this
work, we have investigated in some detail the interplay of thermal
motion, the phase space of decay products, and intrinsic changes in
the structure of the $\rho(776),~\omega(782),~f_0(600),~{\rm
and}~\Delta(1232)$ resonances for temperatures and hadron densities
expected at freeze-out.  The effects of thermal and orbital angular
momentum-dependent phase space functions were evaluated without
approximation.  Analytic results of shifts to leading order in $T/M_R$
($T$ is the temperature and $M_R$ is mass at the peak of the
resonance) for the cases of both equal and unequal mass decay products
were derived.  Medium-induced intrinsic shifts in the width and peak
positions were taken from state-of-the-art calculations in order to
assess the extent to which such shifts are required to account for the
reported shift of the $\rho-$meson peak.  The contribution to the
$2\pi$ final state from the $f_0(600)$ or sigma meson was assessed.
The expected shift in the peak of the invariant mass distribution of
decay of the $\Delta(1232)$ resonance was calculated.

Among the important points that emerge from calculations in which  
resonance peak masses and widths were taken to be at their vacuum values
are: 
\begin{itemize}

\item The combined effects of thermal and decay product phase spaces
on the $\rho-$ peak mass are small, and at best produce a downward
shift of about $10$ MeV (from the vacuum value) at the lowest
temperature in the range $100 \leq T/{\rm MeV} \leq 160$. Upward shifts
occur for temperatures $T >  160$ MeV.

\item The $\sigma-$invariant distribution is considerably distorted,
chiefly because of its large vacuum width.  Depending strongly on the
parameters used to describe its mass spectrum in vacuum, shifts of
hundreds of MeV occur in its peak mass in a much broadened mass
distribution.

\item The shift in the peak mass of the $\omega-$resonance is
negligibly small because of its small vacuum width.

\item Largely due to the growing phase space of the decay products, 
the peak of the $\Delta-$resonance shifts upward from its vacuum
value in the range $100 < T_f/{\rm MeV} < 120$, where
$T_f$ is the freeze-out temperature. Downward shifts are produced only
for (unlikely) freeze-out temperatures $T_f \leq 75$ MeV.     

\end{itemize} 

For freeze-out temperatures in the range $100 < T_f/{\rm MeV} < 125$,
calculations in which intrinsic changes in the resonance widths and
masses were incorporated led to the following conclusions:

\begin{itemize}

\item Although significant downward mass shifts are caused by widened
widths of the $\rho-$meson, a sizeable shift in its intrinsic mass, at
least of order 50 MeV, is required to generate a shift of order 60-70
MeV in the peak of the invariant mass distribution of the
$\rho-$meson.

\item Downward shifts of hundreds of MeV occur in the
$\sigma-$invariant distribution with decreasing in-medium $\sigma$
masses suggested by chiral symmetry restoration.  As a result,
contributions from the $\sigma \rightarrow \pi\pi$ decay at the peak
$\rho-$mass are primarily determined by the exponentially decreasing
thermal phase space and provide small corrections to those from $\rho
\rightarrow \pi\pi$ decays. The summed invariant distribution is
relatively flat in the range $2m_{\pi} < M \leq 4.5m_\pi$. In this
region, however, the $K_0^s(498) \rightarrow \pi\pi$ decays exhibit a
prominent peak at $\sim 500$ MeV and the $\sigma \rightarrow \pi\pi$
decays constitute a background to this peak.

\item Unless large intrinsic shifts are caused by the medium, the
shift of the $\omega-$peak in the invariant distribution of its 
$\pi^+\pi^-$ final state is very small.

\item An observed downward shift from the vacuum peak at 1232
MeV from $\Delta-$decay distinctively signals a downward shift in its
intrinsic peak mass due to in-medium effects.  (Recall that, unlike
for the $\rho-$meson, thermal and decay product phase space functions
produce an upward shift for the $\Delta$.)  For example, at $T_f=100$
MeV, the peak mass in the invariant distribution moves down by
$\sim 10(30)$ MeV for an intrinsic downward mass shift of $20(40)$
MeV.

\end{itemize} 

\se{Outlook }

Some points that are naturally raised as a consequence of this work
and that merit further study include: 

\begin{itemize}

\item The reported shift in the $\rho-$meson peak is surprisingly
large. Why and how then are the particles decoupling insofar as
freeze-out represents the stage when interactions are at their
weakest? The answers to these questions hinge on the extent to which
rescattering effects affect the decay distributions at freeze-out
(see, {\it e.g.,} Ref.~\cite{BBFSB91}).

\item A shift as large as 60-70 MeV in the $\rho-$mass at the dilute
stages points to considerably larger shifts at the dense stages of the
collision. This has far-reaching consequences for dilepton and photon
emissions, which occur predominantly at hadron densities and
temperatures that are significantly larger than those at freeze-out.
Experimental establishment of spectroscopic shifts, both at the dilute
stage (through hadronic probes) and dense stage (through
electromagnetic probes) at RHIC collisions offers the opportunity to
choose between scenarios that have distinctive predictions about the
effects of chiral symmetry restoration on resonances in a hot and
dense medium.

\item Spectroscopy of resonance decays, particularly those of the
$\rho-$meson and $\Delta$-isobar, also offers the means to delimit the
temperature and baryon chemical potential of hadronic matter at the
decoupling stage (for more details, see Secs. II and IV). This
complements our knowledge about freeze-out conditions inferred from
stable particle ratios.

\item We cannot refrain from mentioning that it would be marvelous,
albeit difficult, to conduct resonance tomography through
Hanbury-Brown--Twiss analyses by utilizing the decay products of
resonances at RHIC and planned LHC experiments.

\end{itemize} 

\section*{Acknowledgements}


We thank Gerry Brown, Joe Kapusta, Volker Koch, Ralf Rapp, and Edward
Shuryak for enlightening discussions.  This work was supported in part
by the U.S. Department of Energy under grant No. DE-FG02-88ER40388.
PFK acknowledges support from the Alexander von Humboldt Foundation
through a Feodor-Lynen Fellowship.


\end{narrowtext}

\end{document}